\documentclass[twocolumn,prb]{revtex4}

\usepackage{color}
\usepackage{array}
\usepackage{amsmath}
\usepackage{amssymb}
\usepackage{wasysym}
\usepackage{bm}
\usepackage{graphicx}
\usepackage{txfonts}
\usepackage{epsfig}

\def\bra#1{\left\langle#1\right|}
\def\ket#1{\left|#1\right\rangle}

\def\be{\begin{equation}}       \def\ee{\end{equation}}
\def\bea{\begin{eqnarray}}      \def\eea{\end{eqnarray}}
\def\ba{\begin{array} }
\def\ea{\end{array} }
\def\bnum{\begin{enumerate} }
\def\enum{\end{enumerate}}

\def\=>{\Rightarrow}
\def\>{\rightarrow}

\begin{document}
\title{Identifying non-Abelian topological ordered state and transition by momentum polarization}
\author{Yi Zhang and Xiao-Liang Qi}
\affiliation{Department of Physics, Stanford University, Stanford,
California 94305, USA}
\date{\today}
\begin{abstract}
Using a method called momentum polarization, we study the
quasiparticle topological spin and edge-state chiral central charge
of non-Abelian topological
ordered states described by Gutzwiller-projected wave functions. 
Our results verify that the fractional Chern insulator state
obtained by Gutzwiller projection of two partons in bands of Chern
number $2$ is described by $SU(2)_2$ Chern-Simons theory coupled to
fermions, rather than the pure $SU(2)_2$ Chern-Simons theory. In
addition, by introducing an adiabatic deformation between one Chern
number $2$ band and two Chern number $1$ bands, we show that the
topological order in the Gutzwiller-projected state does not always
agree with the expectation of topological field theory. Even if the
parton mean-field state is adiabatically deformed, the Gutzwiller
projection can introduce a topological phase transition between
Abelian and non-Abelian topologically ordered states. Our approach
applies to more general topologically ordered states described by
Gutzwiller-projected wave functions.
\end{abstract}
\maketitle

\section{Introduction}

Topologically ordered states (TOSs) are unconventional states of
matter with ground-state degeneracy, elementary quasiparticle
excitations with fractional statistics, and long-range quantum
entanglement\cite{wen1990}. The non-Abelian TOSs are a subcategory
of TOSs in which quasiparticles carry nonlocal topological
degeneracy and have received much recent attention due to their
potential applications in topological quantum
computations\cite{memory, kitaev2003, nayak}. The braiding processes
of quasiparticles within a non-Abelian TO induce noncommuting
unitary transformations in the ground-state space instead of merely
incurring a $U(1)$ phase factor as in the Abelian case. Candidates
for non-Abelian TOSs include the $\nu=5/2$ and $\nu=12/5$ fractional
quantum Hall states\cite{tsui1987}, which are proposed to be the
Moore-Read state\cite{MooreRead} and the Read-Rezayi
states\cite{ReadRezayi}.

Unlike conventional states of matter characterized by the symmetries
preserved or those broken spontaneously, TOSs are characterized by
topological properties such as ground-state degeneracy and fusion
and braiding of topological quasiparticles. Except for some exactly
solvable models, most candidate systems for TOSs can be studied only
by numerical methods such as the density-matrix renormalization
group (DMRG)\cite{Jiang2012} and the variational Monte Carlo
method\cite{wen2002}. To determine the topological order in a
numerically studied system, it is essential to develop numerical
probes of topological properties. The search for more efficient and
general numerical methods has attracted much recent attention.
Various methods have been developed to characterize quasiparticle
statistics based on direct calculation of the Berry phase
\cite{wen1990}, explicit braiding of excitations\cite{Bais} and
modular transformation of ground states with minimum entanglement
entropy\cite{smat}. Recently, an additional approach has been
proposed for numerically extracting two topological properties of a
given TOS, the topological spins of quasiparticles $h_a$ and the
edge-state chiral central charge $c$\cite{mompol}. Physically, the
topological spin determines the phase factor $\theta_a=e^{i2\pi
h_a}$ obtained by the system when a quasiparticle spins through
$2\pi$. The chiral central charge of the edge state determines the
thermal current $I_E=\frac{c}{6}T^2$ at temperature
$T$\cite{Affleck1986}. These two quantities are essential in
determining the TOS. The proposal is based on the concept of {\it
momentum polarization} defined for cylindrical systems. For a
cylindrical lattice system with periodic boundary condition along
the $y$ direction, one can define a unitary ``partial translation
operator" $T_y^L$ which translates the lattice sites along the $y$
direction by one lattice constant for all sites that are in the left
half of the system. For a topological ground state $\ket{\Phi_a}$
with quasiparticle type $a$ in the cylinder, the expectation value
of $T_y^L$ is proposed to have the following asymptotic
form\cite{mompol}

\begin{eqnarray}
\lambda_a\equiv \bra{\Phi_a}T_y^L\ket{\Phi_a}\simeq \exp\left[\frac{2\pi
i}{L_y}p_a-\alpha L_y\right]
\end{eqnarray}
where $L_y$ is the number of lattice sites in the $\hat y$
direction, $\alpha$ is a nonuniversal complex constant for the
leading contribution and independent of the specific topological
sector $a$, and remarkably, the fractional part of the momentum
polarization $p_a$ has a universal value $p_a=h_a-\frac{c}{24}$,
which measures the combination of topological spin $h_a$ (modulo
$1$) and central charge $c$ (modulo $24$). Since $T_y^L$ only acts
only on the left half of the system, the momentum polarization is a
quantum entanglement property determined by the reduced density
matrix of the left half of the system. The average value $\lambda_a$
has the merit of being relatively simple to evaluate in comparison
with the previous methods based on entanglement entropy\cite{smat}.
The calculation of the Renyi entanglement entropy involves a swap
operator and requires a minimum of two replicas of the system, while
for momentum polarization the evaluation of $T_y^L$ does not need a
replica so the Hilbert space for Monte Carlo sampling is much
smaller for the same system size. In Ref. \onlinecite{mompol}, the
momentum polarization was studied for two simple TOSs, the Laughlin
1/2 state in fractional Chern insulators (the definition of which
will be given in the next paragraph) and the honeycomb lattice
Kitaev model\cite{kitaev2}. The former is an Abelian state, while
the latter has a special non-Abelian state that can be solved by
mapping to free Majorana fermions.

In this paper, we apply the momentum polarization approach to more
generic non-Abelian TOSs. More specifically, we study non-Abelian
states described by Gutzwiller-projected wave
functions\cite{gros1989} of fractional Chern insulators (FCIs). An
(integer) Chern insulator is a band insulator with nonzero quantized
Hall conductance. The Hall conductance $\sigma_H=n\frac{e^2}{h}$
carried by an occupied band is determined by a topological invariant
of the energy band, known as the Chern number $C=n$. FCIs are
generalizations of Chern insulators to interacting systems, which
have fractional Hall conductance and topological order. One way to
understand FCIs is through the parton construction, in which the
electron is considered as a composite particle of several ``partons"
carrying fractional quantum numbers. For example, an electron can be
split into three fermionic partons, with each parton in an integer
Chern insulator with $C=1$. The corresponding electron state has
Hall conductance $\frac13\frac{e^2}{h}$ and is the $\frac13$
Laughlin state. Gauge fields are coupled to partons to enforce the
constraint that all physical states are electron states and no
individual parton will be observed. The parton construction can be
expressed in ansatz ground state wave functions constructed by the
procedure of Gutzwiller projection \cite{gros1989}, which is a
projection of the parton ground state into the physical electron
Hilbert space. Gutzwiller-projected wave functions have been
constructed for FCI\cite{frank2011b}. When two partons are glued
together to form a bosonic ``electron", and each parton is in a
state with Chern number $C=1$, from topological effective field
theory (which we will review later in the paper) one expects to find
a $1/2$ bosonic Laughlin state. In contrast, if each parton is in a
state with Chern number $C=2$, the resulting electron TOS is
expected to be non-Abelian, related to $SU(2)$ level-$2$
Chern-Simons (CS) theory \cite{MooreRead}. The non-Abelian nature of
this state has been verified by calculation of the modular $\mathcal
S$ matrix for the projected wave functions\cite{frank2013}.

In this paper, we study the momentum polarization of the
Gutzwiller-projected wave function for the state of two partons with
Chern number $C=2$. In addition to confirming the non-Abelian
topological order of this state, our result contains the following
two points. First, the spin and central charge obtained from
momentum polarization clearly distinguish two related but distinct
topological states, the $SU(2)_{2}$ CS theory and the $SU(2)_{2}$ CS
theory coupled to fermions\cite{Maissam2010}. The particle fusion,
braiding, and modular $\mathcal S$ matrix of these two theories are
identical, but they are distinct TOSs with different edge-state
chiral central charge $c=\frac{3}2$ and $c=\frac{5}2$, respectively.
The momentum polarization calculation clearly demonstrates that the
Gutzwiller-projected parton wave function has the topological order
of the latter theory. Second, there is an apparent paradox in the
statement that Gutzwiller projection of parton $C=2$ states leads to
$SU(2)_2$ CS theory coupled to fermions. Since Chern number is the
only topological invariant of a fermion energy band, a Chern number
$C=2$ band can be adiabatically deformed to two decoupled $C=1$
bands, as long as translation symmetry breaking is allowed. Since
the Gutzwiller projection of two $C=1$ partons is known to give the
Laughlin $1/2$ state, it appears that one can adiabatically deform
the non-Abelian TOS obtained from partons occupying the $C=2$ band
to the Abelian TOS of two decoupled Laughlin $1/2$ states. This is
clearly in contradiction with the topological stability of TOSs. By
introducing an explicit adiabatic deformation between a $C=2$ band
structure and two decoupled $C=1$ bands, we study the quasiparticle
topological spin during the adiabatic interpolation. Our result
shows that there is a topological phase transition between the
Abelian phase of the bilayer Laughlin state and the non-Abelian
phase of the $SU(2)_2$ CS coupled to fermions. The topological phase
transition occurs at a {\it finite} coupling between the two $C=1$
bands. In other words, the TOS obtained from Gutzwiller projection
of $C=2$ parton bands is {\it not} completely determined by the
Chern number of the parton band structure, but may depend on details
of the Chern bands and the projection. The argument based on parton
``mean-field theory", {\it i.e.}, integrating over partons to obtain
CS gauge theory, may not predict the correct phase. This example
further emphasizes the importance of numerical approaches such as
momentum polarization in identifying TOSs. Based on this numerical
observation, we will also discuss theoretically the effective theory
interpretation of this topological phase transition.

The remaining of the paper is organized as follows: In Sec.
\ref{sec2}, we present our momentum polarization calculation in the
Gutzwiller-projected wave function of non-Abelian FCIs, after
reviewing the relevant background knowledge. Sec. \ref{sec2A}
presents our projective construction and the $C=2$ Chern insulator
model; Sec. \ref{sec2B} gives a brief field theory discussion of the
corresponding TOS; Sec. \ref{sec2C} shows our numerical results from
momentum polarization. We obtain the topological spin of the
non-Abelian quasiparticle $h_\sigma=0.321\pm0.013$ and the fermion
quasiparticle $h_\psi=0.520\pm0.026$ and edge central charge
$c=2.870\pm0.176$, in agreement with the $SU(2)$ CS theory coupled
to fermions ($h_\sigma=5/16$ and $c=5/2$). In Sec. \ref{sec3}, we
introduce the adiabatic deformation between two $C=1$ bands and one
$C=2$ band, and study the topological phase transition between the
two TOSs. In Sec. \ref{sec3A}, we present an adiabatic interpolation
of the parton tight-binding Hamiltonian. Sec. \ref{sec3B} presents
the results for the quasiparticle topological spin and ground-state
degeneracy which indicate the transition between the non-Abelian and
Abelian TOS; In Sec. \ref{sec3C}, we discuss the physical
interpretation of this topological transition. Finally, Sec
\ref{sec4} is devoted to a conclusion from our main results and
discussion of open questions.

\section{Identifying the non-Abelian TOS in $C=2$ FCI} \label{sec2}

\subsection{The projective construction and $C=2$ Chern insulator model}  \label{sec2A}

The projective construction is a powerful formalism for ansatz wave
functions of many TOS\cite{wen19913}. For our projective
construction, we first introduce several species of partons $\psi_a$
as free fermions in a Chern insulator, and then constrain the
partons to recombine into physical ``electrons" (which may be bosons
or fermions). In the simple Gutzwiller-projected states we will
discuss in this work,  the projected wave function is defined in
first quantized language  by
$\Phi\left(\left\{z_i\right\}\right)=\underset{a}{\prod}\psi_a\left(\left\{z_i\right\}\right)$.
Here $\left\{z_i\right\}$ with $i=1,2,....,N$ are the coordinates of
all particles, and $\psi_a\left(\left\{z_i\right\}\right)$ is the
wave function of the $a$-th parton. $N$ is the number of each parton
type, which is the same as the total electron number of the system.
The properties of the resulting states can be numerically computed
through variational Monte Carlo calculations.

\begin{figure}
\begin{center}
\includegraphics[scale=0.3]{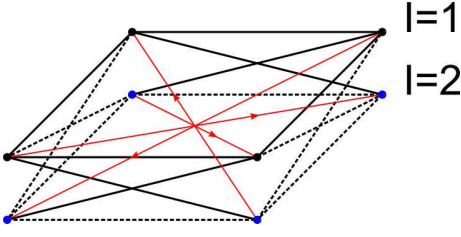}
\caption{An illustration of the hopping Hamiltonian in Eq.
\ref{HamC2}. The two orbitals on each lattice site are shown as
different layers and colored in black and blue, respectively. The
hopping is $+1$ ($-1$) along the solid (dashed) lines, and $i/\sqrt
2$ ($-i/\sqrt 2$) along (against) the red arrows.} \label{fig7}
\end{center}
\end{figure}

For our focused non-Abelian TOS, we start with the following parton
mean-field Hamiltonian on a two-dimensional square lattice
\begin{eqnarray}
H_{C=2}&=&\underset{<ij>,I,s}{\sum}(-1)^{I-1}c_{jIs}^{\dagger}c_{iIs}+\underset{<ij>,s}{\sum}e^{i2\theta_{ij}}\left(c_{j2s}^{\dagger}c_{i1s}+c_{j1s}^{\dagger}c_{i2s}\right)
\nonumber \\
&+&\frac{1}{\sqrt{2}}\underset{<<ik>>,s}{\sum}e^{i2\theta_{ik}}\left(c_{k2s}^{\dagger}c_{i1s}-c_{k1s}^{\dagger}c_{i2s}\right)+\mbox{H.C.}\nonumber\\
&=&\underset{<ij>_{y},s}{\sum}\left[\left(c_{j1s}^{\dagger}c_{i1s}-c_{j2s}^{\dagger}c_{i2s}\right)-\left(c_{j2s}^{\dagger}c_{i1s}+c_{j1s}^{\dagger}c_{i2s}\right)\right]
\nonumber \\
&+&\underset{<ij>_{x},s}{\sum}\left[\left(c_{j1s}^{\dagger}c_{i1s}-c_{j2s}^{\dagger}c_{i2s}\right)+\left(c_{j2s}^{\dagger}c_{i1s}+c_{j1s}^{\dagger}c_{i2s}\right)\right]
\nonumber \\
&+&\frac{1}{\sqrt
2}\underset{<<ik>>,s}{\sum}e^{i2\theta_{ik}}\left(c_{k2s}^{\dagger}c_{i1s}-c_{k1s}^{\dagger}c_{i2s}\right)+\mbox{H.C.}
\label{HamC2}
\end{eqnarray}
where $I=1,2$ are the two orbitals on each lattice site and
$s=\uparrow, \downarrow$ labels the two flavors of partons.
$\theta_{ij}$ is the azimuthal angle for the vector connecting $i$
and $j$. $<ij>$ and $<<ik>>$ label nearest neighbor and next nearest
neighbor links, while $\left\langle ij\right\rangle _{x}$ and
$\left\langle ij\right\rangle _{y}$ denote nearest neighbors along
the $\hat x$ and $\hat y$ directions, respectively, as is
illustrated in Fig. \ref{fig7}. A previous study\cite{frank2013} has
shown that at half filling the system is a Chern insulator with
$C=2$. The correlation length $\xi$ is on the order of a lattice
constant, and therefore the finite-size effects are generally
suppressed for the system sizes we study.

In real space, the parton wave function $\psi_a
\left(\left\{z_{i}\right\}\right)$ is a Slater determinant for a
completely filled valence band, where $z=(i, I)$ labels both the
position and orbital indices of a parton. Next we apply the
Gutzwiller projection imposing the constraint
$n_{iI\uparrow}=n_{iI\downarrow}$, with $n_{iIs}=c_{iIs}^\dagger
c_{iIs}$ the number of partons at each site and orbital. The states
satisfying this constraint have two partons bound at each site and
orbital, and are physical electron states with electron number
$n_{iI}^e=n_{iI\uparrow}$. This leads to the following many-body
wave function
\begin{eqnarray}
\Phi\left(\left\{z_{i}\right\}\right)=\psi_{\uparrow}\left(\left\{z_{i}\right\}\right)\psi_{\downarrow}\left(\left\{z_{i}\right\}\right)
=\psi^2_{\uparrow}\left(\left\{z_{i}\right\}\right)
\end{eqnarray}
This state is the major focus of the paper. Previously, the three
topological sectors on a torus for this projective construction were
obtained by tuning the boundary condition of the parton mean-field
Hamiltonian in Eq. \ref{HamC2} and their connection to the
corresponding threaded quasiparticle has been
established\cite{frank2013}. For our momentum polarization
calculations, we need to generalize the projective construction to a
cylinder. To resolve the complication from the gapless chiral edge
modes on the open edges, we start from a torus and adiabatically
lower all hopping amplitudes across the open boundary until they are
much smaller than the edge modes' finite size gap. The residue
hoppings effectively couple only the zero energy states at $k_y=\pm
\pi/2$ on the two edges of the cylinder, therefore the original
boundary conditions of topological sectors on the torus lead to
linear combinations of the zero energy states\cite{mompol}. Since
such a process involves no level crossing, we can obtain the
topological sectors on a cylinder by allowing occupation of
different parton zero-energy states on the two edges.

\subsection{Topological Field Theory Description} \label{sec2B}

To understand the TOS described by the above projective
construction, we briefly review the topological field theory
description of this state. The electron operator can be expressed in
partons as $f_{iI}=c_{iI\uparrow}c_{iI\downarrow}$. This
decomposition has an $SU(2)$ gauge symmetry: for any $SU(2)$ matrix
with $\alpha,\beta \in \mathbb{C}$ and
$\left|\alpha\right|^2+\left|\beta\right|^2=1$
\begin{eqnarray}
\left(\begin{array}{c}
c_{iI\uparrow}\\
c_{iI\downarrow}
\end{array}\right)\rightarrow\left(\begin{array}{cc}
\alpha & \beta\\
-\beta^{*} & \alpha^{*}
\end{array}\right)\left(\begin{array}{c}
c_{iI\uparrow}\\
c_{iI\downarrow}
\end{array}\right)
\end{eqnarray}
this transformation preserves the electron operator
$f_{iI}\rightarrow\left(\alpha c_{iI\uparrow}+\beta
c_{iI\downarrow}\right)\left(-\beta^{*}c_{iI\uparrow}+\alpha^{*}c_{iI\downarrow}\right)=c_{iI\uparrow}c_{iI\downarrow}=f_{iI}$,
and therefore the effective theory of partons should also be gauge
invariant. The simplest possible effective theory satisfying the
gauge invariant condition is obtained by a minimal coupling of the
mean-field Hamiltonian (\ref{HamC2}) to an $SU(2)$ gauge
field\cite{wen1999}. A lattice $SU(2)$ gauge field is described by
gauge connection $e^{ia_{ij}}\in SU(2)$ defined along each link
$ij$.  The Hamiltonian is written as
\begin{eqnarray}
H_{eff}&=&\underset{<ij>,I}{\sum}(-1)^{I-1}e^{ia^{ji}_{sr}}c_{jIs}^{\dagger}c_{iIr}+\underset{<ij>}{\sum}e^{i2\theta_{ij}}e^{ia^{ji}_{sr}}\left(c_{j2s}^{\dagger}c_{i1r}+c_{j1s}^{\dagger}c_{i2r}\right)
\nonumber \\
&+&\frac{1}{\sqrt{2}}\underset{<<ik>>}{\sum}e^{i2\theta_{ik}}e^{ia^{ki}_{sr}}\left(c_{k2s}^{\dagger}c_{i1r}-c_{k1s}^{\dagger}c_{i2r}\right)+\mbox{H.C.}
\label{Hameff}
\end{eqnarray}
where $s,r=\uparrow,\downarrow$ denote the two parton species, and
repeated indices are summed over.

Since the partons are gapped, it is straightforward to integrate
them out. Due to the Chern number $C=2$ of each parton band,
integrating over the parton results in an $SU(2)_2$ non-Abelian CS
theory
\begin{eqnarray}
\mathcal L= \frac{2}{4\pi} \epsilon_{\mu\nu\rho} \mbox{tr}
\left[a_\mu\partial_\nu a_\rho+\frac{2}{3} a_\mu a_\nu a_\rho\right]
\label{LSU22}
\end{eqnarray}

However, it is not accurate to say that the topological field theory
describing the TOS of this parton construction is $SU(2)_2$ CS gauge
theory, because the partons have nontrivial contribution to
topological properties such as edge theory. The edge theory of
$SU(2)_2$ CS theory is a chiral $SU(2)_2$ Weiss-Zumino-Witten (WZW)
model\cite{wess1971,witten1983}, while the edge theory of the FCI
described above consists of four chiral fermions (two from each
flavor of parton) coupled to the $SU(2)_2$ WZW model. Technically,
the edge state of fermions coupled to the WZW model is described by
a quotient of two conformal field theories $\frac{U(4)_1}{SU(2)_2}$,
in which $U(4)_1$ describes four free chiral fermions and $SU(2)_2$
describes the gauge degrees of freedom which are removed from
physical excitations.\cite{Maissam2010} Although they both have
three quasiparticles with the same fusion rule and braiding
statistics, these two theories are not topologically equivalent. In
particular, the topological spin differs by a fermionic sign for
quasiparticles which correspond to an odd number of holes in the
parton Chern insulator state. For comparison purpose, we list the
theoretical values for the quasiparticle topological spins and edge
central charges for the two theories in Table \ref{table1}.

\begin{table}
\begin{center}
\begin{tabular}{|c|c|c|}
\hline
 & $SU(2)_{2}$ CS & $\nu=2$ coupled to $SU(2)_2$\tabularnewline
\hline \hline $c$ & 3/2 & 5/2\tabularnewline \hline $h_{1}$ & 0 &
0\tabularnewline \hline $h_{\sigma}$ & 3/16 & 5/16\tabularnewline
\hline $h_{\psi}$ & 1/2 & 1/2\tabularnewline \hline $D$ & 3 &
3\tabularnewline \hline
\end{tabular}
\caption{Theoretical values of topological properties including the
edge central charge $c$, the topological spins for quasiparticles
$h_1$, $h_\sigma$, $h_\psi$ and the ground-state degeneracy $D$ for
the pure $SU(2)_2$ CS theory and the $\nu=2$ fermions coupled to an
$SU(2)_2$ gauge field (or equivalently, $\frac{U(4)_1}{SU(2)_2}$
theory).} \label{table1}
\end{center}
\end{table}

In summary, we have seen that the effective topological field theory
analysis suggests that the topological order in the
Gutzwiller-projected state is $\frac{U(4)_1}{SU(2)_2}$ instead of
$SU(2)_2$. However, it is essential to verify that directly for the
Gutzwiller-projected wave function, as there is no guarantee that
the effect of Gutzwiller projection is completely equivalent to the
coupling to a gauge field in the effective field theory. This is
achieved in the next section by studying the momentum polarization.

\subsection{Topological spin and edge central charge from momentum polarization calculations} \label{sec2C}

\begin{figure}
\begin{center}
\includegraphics[scale=0.4]{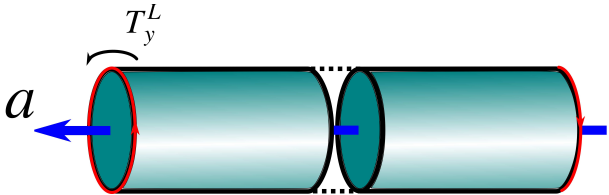}
\caption{The partial translation operator $T^L_y$ translates the
left half of the cylinder by one lattice constant along the $\hat y$
direction. The red arrows indicate the chiral edge modes. The
topological sector $a$ is determined by the type of quasiparticle
threaded through the cylinder, denoted by the large blue arrow.}
\label{fig3}
\end{center}
\end{figure}

Quasiparticle braiding from previous studies has determined that the
TOS for $\Phi\left(\left\{z_i\right\}\right)$ is necessarily
non-Abelian. However, both theories in Table 1 are consistent with
the braiding, and therefore additional information is necessary to
make a complete identification. We numerically extract the
quasiparticle topological spin and edge central charge from momentum
polarization calculations for the model in Eq. \ref{HamC2} defined
on a cylinder.

Care should be taken about the non-Abelian topological sector, which
consists of parton states with an overall difference of momentum
$\pi$ on the left edge. For the expectation value of the partial
translation operator $T^L_y$ that translates the left half of the
cylinder by one lattice constant along the $\hat y$ direction, see
Fig. \ref{fig3} for illustration, this $\pi$ momentum difference
will result in contributions with opposite signs. To overcome this
difficulty, we generalize $T^L_y$ to twist the left half of the
cylinder by $l$ lattice constants, so that the overall phase
difference vanishes for a partial translation of $l=2$ lattice
constants. For this purpose, we take $\frac{L_y}{l}$ to be integer,
consider $l$ sites along the $\hat{y}$ direction as one unit cell,
and replace $L_y$ by $\frac{L_y}{l}$ in the formula proposed in Ref.
\onlinecite{mompol}. Consequently, the average value of $T_y^L$
defined by $\lambda_{a}=\bra{\Phi_{a}}T_{y}^{L}\ket{\Phi_{a}}$ has
the following leading contributions
\begin{eqnarray}
\lambda_{a}=\exp\left[i\frac{2\pi l}{L_{y}}p_a-\alpha
\frac{L_{y}}l\right] \label{Mompol}
\end{eqnarray}
in which $\alpha$ is a nonuniversal complex constant independent of
the specific topological sector $a$, while $p_a$ has a universal
topological value $p_a=h_{a}-c/24$ determined by the topological
spin $h_{a}$ and the edge central charge $c$.

The quantity in Eq. \ref{Mompol} can be efficiently evaluated for
the projected wave functions with the variational Monte Carlo
method. For a cylinder with $L_{x}=8$, $L_{y}=16$ and $T_{y}^{L}$
translating the left half by $l=2$ lattice constants for the
aforementioned reason, numerical calculations yield
$\arg\left(\lambda_{1}\right)=-3.4449\pm0.0063$ for the identity
sector, $\arg\left(\lambda_{\sigma}\right)=-3.1929\pm0.0082$ for the
sector associated with the non-Abelian quasiparticle, and
$\arg\left(\lambda_{\psi}\right)=-3.0366 \pm 0.0257$ for the fermion
sector. With $h_1=0$ by definition of the identity particle, we
obtain
\begin{eqnarray}
h_{\sigma}=\frac{L_{y}}{2\pi
l}\left[\arg\left(\lambda_{\sigma}\right)-\arg\left(\lambda_{1}\right)\right]=0.321\pm0.013
\\
h_{\psi}=\frac{L_{y}}{2\pi
l}\left[\arg\left(\lambda_{\psi}\right)-\arg\left(\lambda_{1}\right)\right]=0.520\pm0.026
 \label{hsigma}
\end{eqnarray}

This is fully consistent with the theoretical value of
$h_{\sigma}^{th}=5/16=0.3125$ for the non-Abelian quasiparticle and
$h_{\psi}^{th}=1/2=0.5$ for the fermion quasiparticle of a theory of
$\nu=2$ fermions couple to an $SU(2)$ gauge field .

\begin{figure}
\begin{center}
\includegraphics[scale=0.4]{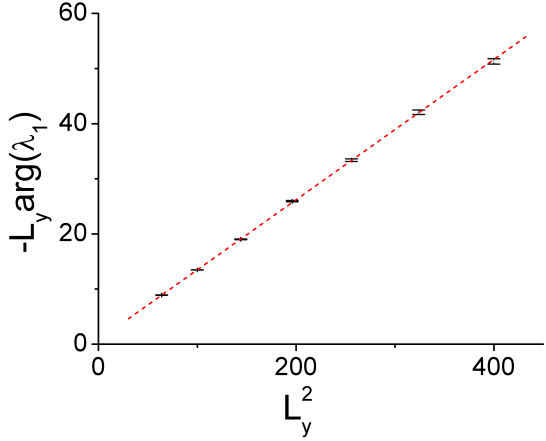}
\caption{The value of $-L_y \arg \left(\lambda_1\right)$ versus
$L^2_y$ for the identity sector $a=1$. The intercept at $L^2_y=0$ of
the linear fitting gives $-2\pi p_1=0.7513\pm 0.046$. We set $L_x=8$
and $l=1$ for all calculations.} \label{fig1}
\end{center}
\end{figure}

In addition, we calculate $\lambda_{1}$ for $L_{x}=8$, $l=1$, and
various values of $L_{y}$. The numerical results are shown in Fig.
\ref{fig1}. To compare with Eq. \ref{Mompol}, note that
$-L_{y}\arg\left(\lambda_{1}\right)=\mbox{Im}\alpha L_{y}^{2}-2\pi
p_1$, so the intercept of this linear fitting gives the value of
$-2\pi p_1=2\pi c/24=0.7513\pm0.046$. The resulting value of
$c=2.870\pm0.176$ is also fairly consistent with the prediction of
$c^{th}=5/2$ according to the theory of $\nu=2$ fermions coupled to
an $SU(2)$ gauge field. Although there is a deviation between the
numerical value and the theoretical value $5/2$ which is probably
due to the finite-size effect, the accuracy of the result is
sufficient to completely distinguish this system from the bare
$SU(2)_{2}$ CS theory with $h_\sigma=3/16$ and $c=3/2$. This result
also provides further evidence that the momentum polarization method
for computing topological quantities is applicable to non-Abelian
TOSs.

\section{The transition between Abelian and non-Abelian TOS in projected wave functions} \label{sec3}

From the results discussed in the last section, it seems that the
TOS of the Gutzwiller-projected wave function agrees well with the
expectation from the topological field theory approach. However,
there is a hidden paradox in this result. Since the Chern number is
the only topological invariant for a generic energy band in two
dimensions, a band with Chern number $C=2$ is topologically
equivalent to two $C=1$ bands. More explicitly, an exact mapping has
been constructed between a $C=2$ band and two decoupled Landau level
systems which are related by a lattice translation
operation\cite{qi2011b,Maissam2012,wu2013}. Therefore one would
naively expect that a state with each parton in a $C=2$ Chern
insulator is adiabatically equivalent to one in which each parton
occupies two  $C=1$ bands. However, this statement seems to
contradict the fact that the Gutzwiller-projected wave function of
the latter state is Abelian. It is known that the
Gutzwiller-projected wave function of two partons each in a $C=1$
band gives a Laughlin $\nu=\frac12$ Abelian TOS\cite{kalmeyer,
kalmeyer1989, wenzee, frank2011b, smat, mompol}, which is also
denoted $SU(2)_1$ Chern-Simons theory. Therefore one would expect
that when each parton occupies two decoupled $C=1$ bands, which can
be viewed as two decoupled layers, the Gutzwiller-projected wave
function of the whole system is simply two copies of the Laughlin
$\nu=\frac12$ state, {\it i.e.} $SU(2)_1\times SU(2)_1$, which is an
Abelian state clearly distinct from the $\frac{U(4)_1}{SU(2)_2}$
theory we obtained earlier from both effective theory and numerical
results. To resolve this apparent paradox, in this section we
introduce an explicit interpolation between the $C=2$ model used in
last section and a model with two decoupled $C=1$ bands. By studying
the momentum polarization of the corresponding Gutzwiller-projected
wave functions during this interpolation, we find a topological
phase transition between the Abelian and non-Abelian phases.

\subsection{An adiabatic interpolation of the parent Hamiltonian} \label{sec3A}

\begin{figure}
\begin{centering}
\includegraphics[scale=0.4]{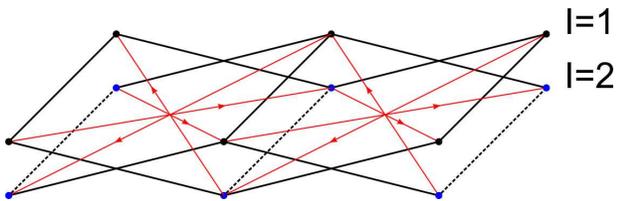}
\par\end{centering}
\caption{An illustration of the hopping Hamiltonian in Eq.
\ref{Ham0}. The two orbitals on each lattice site are shown in
different layers and colored in black and blue, respectively. The
hoppings along the solid (dashed) lines are $+\sqrt 2$ ($-\sqrt 2$),
and along (against) the red arrows are $i/\sqrt 2$ ($-i/\sqrt 2$).
It is straightforward to separate the system into two uncoupled
zigzag subsystems with odd and even values of $x_i+I$.} \label{fig8}
\end{figure}

\begin{figure}
\begin{centering}
\includegraphics[scale=0.3]{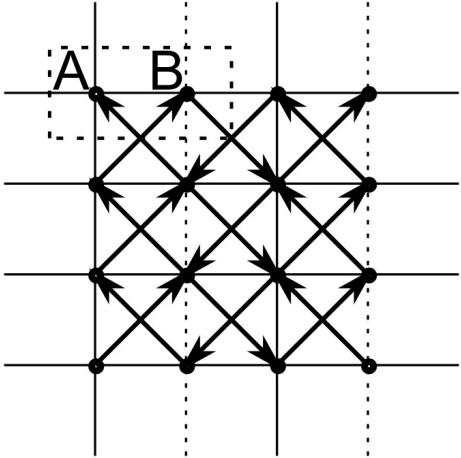}
\par\end{centering}
\caption{Illustration of a $C=1$ Chern insulator model on a
two-dimensional square lattice. The nearest neighbor hopping
amplitudes are $\sqrt 2$ along the square edges and $-\sqrt 2$ along
the dashed lines. The next nearest neighbor hoppings are along the
square diagonal with amplitude $+i/\sqrt 2$  along ($-i/\sqrt 2$
against) the arrow. The two lattice sites in the unit cell are
marked as $A$ and $B$.} \label{fig4}
\end{figure}

As an explicit example of the interpolation between a $C=2$ band and
two $C=1$ bands, we consider the following parton mean-field
Hamiltonian on a two-dimensional square lattice\cite{lee2013}
\begin{eqnarray}
H_\Theta &=&\sqrt
2\underset{<ij>_{y},s}{\sum}\left[\cos\Theta\left(c_{j1s}^{\dagger}c_{i1s}-c_{j2s}^{\dagger}c_{i2s}\right)-\sin\Theta\left(c_{j2s}^{\dagger}c_{i1s}+c_{j1s}^{\dagger}c_{i2s}\right)\right]
\nonumber \\
&+&\sqrt
2\underset{<ij>_{x},s}{\sum}\left[\sin\Theta\left(c_{j1s}^{\dagger}c_{i1s}-c_{j2s}^{\dagger}c_{i2s}\right)+\cos\Theta\left(c_{j2s}^{\dagger}c_{i1s}+c_{j1s}^{\dagger}c_{i2s}\right)\right]
\nonumber \\
&+&\frac{1}{\sqrt
2}\underset{<<ik>>,s}{\sum}e^{i2\theta_{ik}}\left(c_{k2s}^{\dagger}c_{i1s}-c_{k1s}^{\dagger}c_{i2s}\right)+\mbox{H.C.}
\label{HamTheta}
\end{eqnarray}
where the label definition is the same as in Eq. \ref{HamC2}, and
$\Theta$ is a continuous parameter. For $\Theta=\pi/4$, Eq.
\ref{HamTheta} returns to the Hamiltonian in Eq. \ref{HamC2} with a
$C=2$ band. For $\Theta=0$, the Hamiltonian becomes
\begin{eqnarray}
H_{\Theta=0} &=&\sqrt
2\underset{<ij>_{y},s}{\sum}\left(c_{j1s}^{\dagger}c_{i1s}-c_{j2s}^{\dagger}c_{i2s}\right)
+\sqrt 2
\underset{<ij>_{x},s}{\sum}\left(c_{j2s}^{\dagger}c_{i1s}+c_{j1s}^{\dagger}c_{i2s}\right)
\nonumber \\
&+&\frac{1}{\sqrt
2}\underset{<<ik>>,s}{\sum}e^{i2\theta_{ik}}\left(c_{k2s}^{\dagger}c_{i1s}-c_{k1s}^{\dagger}c_{i2s}\right)+\mbox{H.C.}
\label{Ham0}
\end{eqnarray}
The hopping matrix elements are drawn in Fig. \ref{fig8}. Since
hoppings exist only between $I=1$($I=2$) orbitals on the $x_i$ odd
sites and $I=2$($I=1$) orbitals on the $x_i$ even sites, the system
can be directly decomposed into two uncoupled subsystems with even
and odd values of $x_i+I$. The two subsystems are related by a
translation by one lattice constant along the $\hat{x}$ direction.
Suppressing the orbital index, each of the two subsystems has the
following Hamiltonian, which is a Chern insulator with $C=1$ for
each parton flavor $s$
\begin{eqnarray}
H_{C=1}=\underset{\left\langle ij\right\rangle,s
}{\sum}t_{i,j}c_{is}^{\dagger}c_{js}+ \underset{\left\langle
\left\langle ik\right\rangle \right\rangle,s
}{\sum}\Delta_{i,k}c_{is}^{\dagger}c_{ks} + \mbox{H.C.}
\label{HamC1}
\end{eqnarray}
where the nearest neighbor hopping amplitude $t_{i,j}$ is $\sqrt 2$
along the $\hat x$ direction and alternates between $\sqrt 2$ and
$-\sqrt 2$ along the $\hat y$ direction, and the next nearest
neighbor is $\Delta_{i,k} = i/\sqrt 2$ along the arrow and
$\Delta_{i,k} = -i/\sqrt 2$ against the arrow, see Fig. \ref{fig4}
for an illustration. The unit cell contains two lattice sites.
Therefore, Eq. \ref{HamTheta} defines an interpolation between one
Chern insulator with $C=2$ and two decoupled Chern insulators each
with $C=1$.

It it also verified that the interpolation is adiabatic and the band
gap remains finite for all $\Theta$. Actually, the Hamiltonians with
different $\Theta$ can be related by a global unitary transformation
on the orbital space
\begin{eqnarray}
H_\Theta&=&U^{-1}H_0U\nonumber\\
U&=&\exp\underset{s}{\sum}\left[\frac{\Theta}2\left(c_{i1s}^\dagger
c_{i2s}-c_{i2s}^\dagger c_{i1s}\right)\right]\label{unitrans}
\end{eqnarray}
The effect of the rotation on annihilation operators is
\begin{eqnarray}
U^{-1}\left(\begin{array}{c}c_{i1s}\\c_{i2s}\end{array}\right)U&=&\left(\begin{array}{cc}\cos\frac{\Theta}{2}&-\sin\frac{\Theta}{2}\\
\sin\frac{\Theta}{2}&\cos\frac{\Theta}{2}\end{array}\right)\left(\begin{array}{c}c_{i1s}\\c_{i2s}\end{array}\right)
\end{eqnarray}
Consequently, the dispersion and band gap are intact with respect to
the variation of $\Theta$.

\begin{table}
\begin{center}
\begin{tabular}{|c|c|c|}
\hline
 & $SU(2)_1\times SU(2)_1$ CS & $\nu=2$ coupled to $SU(2)_2$\tabularnewline
\hline \hline $c$ & 2 & 5/2\tabularnewline \hline $D$ & 4 &
3\tabularnewline \hline $h$ & 0,1/4,1/4,1/2 &
0,5/16,1/2\tabularnewline \hline
\end{tabular}
\caption{Theoretical values of topological properties including the
edge central charge $c$, ground-state degeneracy $D$, and
quasiparticle topological spins for the $SU(2)_1\times SU(2)_1$ CS
theory and the $\nu=2$ fermions coupled to an $SU(2)_2$ gauge
field.}\label{table2}
\end{center}
\end{table}

Now we study the Gutzwiller-projected state corresponding to the
parton mean-field Hamiltonian $H_\Theta$. We have shown that
$H_{\Theta=\frac{\pi}4}$ leads to the $\frac{U(4)_1}{SU(2)_2}$
state. On the other hand, $H_{\Theta=0}$ describes two decoupled
``layers", each with two partons in $C=1$ bands. The Gutzwiller
projection also applies separately to the two layers, so that the
resulting state is a decoupled bilayer of the projected $C=1$
states. The projected wave functions from a Chern insulator with
$C=1$ have been confirmed to be consistent with the $SU(2)_{1}$ CS
theory\cite{kalmeyer, kalmeyer1989, wenzee, frank2011b, smat,
mompol}. Correspondingly, the projected wave function of two
uncoupled Chern insulators each with $C=1$ should be describable by
an Abelian $SU(2)_{1}\times SU(2)_{1}$ CS theory, which has four
Abelian particles and is clearly distinct from the non-Abelian TOS
established for $\Theta=\frac{\pi}4$. There are major differences in
their topological properties including the torus ground-state
degeneracy, edge central charge and quasiparticle topological spins,
as listed in Table \ref{table2}. Due to this topological difference
between $\Theta=0$ and $\Theta=\frac{\pi}4$, a topological phase
transition must occur for some intermediate $\Theta$. Since the
parton ground states before Gutzwiller projection with different
$\Theta$ are related by a local unitary transformation, one has to
conclude that the topological phase transition is introduced by the
Gutzwiller projection procedure. We study this topological phase
transition numerically in the next section.

\subsection{The quasiparticle topological spin as a signature for topological phase transition} \label{sec3B}

First of all, we would like to determine whether there is a
first-order phase transition at some $\Theta$. Even though the
interpolation of the parton ground state before projection is
clearly adiabatic, the same is not necessarily true for the
projected wave function. Numerically, for $H_\Theta$ defined on a
system of size $L_x=L_y=12$ with periodic boundary conditions, we
study the evolution of the projected wave functions with steps of
$\Theta$ as small as $\delta \Theta = \frac{\pi}{400}$. Variational
Monte Carlo calculations\cite{frank2011b} indicate that for all
values of $\Theta \in [0, \frac{\pi}4]$, the overlap between
neighboring steps' wave functions
$\left|\left\langle\Phi(\Theta+\delta\Theta)|\Phi(\Theta)\right\rangle\right|=1-O\left(10^{-3}\right)$,
which clearly suggests that
$\left\langle\delta\Phi(\Theta)|\Phi(\Theta)\right\rangle
\rightarrow 0$ for small $\delta \Theta \rightarrow 0$ and excludes
the presence of singularities. Therefore the quantum phase
transition must be continuous..

\begin{figure}
\begin{center}
\includegraphics[scale=0.4]{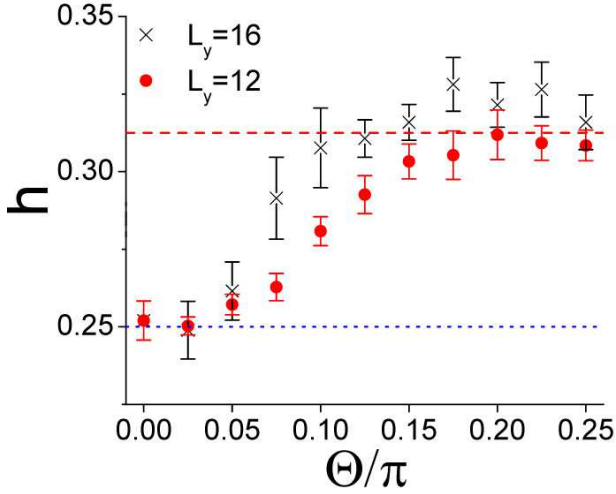}
\caption{The topological spin $h$ for the semion (non-Abelian
quasiparticle) sector versus various values of $\Theta\in[0,\pi/4]$
for the projected Chern insulator in Eq. \ref{HamTheta} from
momentum polarization calculations. The red dashed line and the blue
dotted line are the theoretical values of $h$ for the
$SU(2)_{1}\times SU(2)_{1}$ CS theory ($h_s=1/4$) and $\nu=2$
fermions coupled to an $SU(2)$ gauge field ($h_\sigma=5/16$),
respectively.} \label{fig2}
\end{center}
\end{figure}

In particular, the open boundary conditions are equivalent for the
semion sector in the Abelian TOS and the non-Abelian quasiparticle
sector in the non-Abelian TOSs, as well as for the identity sectors
in both TOS, making an adiabatic interpolation possible within each
sector. To determine the topological phase transition point, we
compute the momentum polarization with $l=2$ for the identity and
semion (non-Abelian quasiparticle) sectors of the projected wave
functions on a cylinder of $L_{x}=8$ and $L_{y}=12, 16$ for each
interpolation of Eq. \ref{HamTheta}. The results of topological spin
$h$ for the semion (non-Abelian quasiparticle) sector versus
$\Theta\in[0,\pi/4]$ are shown in Fig. \ref{fig2}. For small value
of $\Theta=0.05\pi$, the topological spin starts to deviate from the
semionic statistics of $h_s=1/4$ for the Abelian TOS and evolve
towards $h_\sigma=5/16$ for the non-Abelian TOS, see Table
\ref{table2}. Still, there is a finite region of $\Theta$ where the
value of $h$ represents an Abelian TOS. For further verification,
for a smaller value of $\Theta=0.025\pi$, we numerically calculated
the overlaps between projected wave functions of various boundary
conditions on an $L_x=L_y=12$ torus\cite{frank2011b} and find that
there are four linearly independent candidate ground-state wave
functions by projective construction, consistent with the Abelian
$SU(2)_{1}\times SU(2)_{1}$ CS theory. In contrast, for values such
as $\Theta = \pi/4$ and $\Theta = 3\pi/8$ fully in the parameter
region of the non-Abelian topological order, such linear
independence is only three fold.

Our numerical results show that a topological phase transition
occurs at finite $\Theta$, which is consistent with the fact that
the $\Theta=0$ Abelian state is topologically stable and should
persist for a finite region of $\Theta$: the fractional Chern
insulator is an intrinsic topological ordered state protected by an
excitation gap that is stable against small local perturbations of
arbitrary form such as weak couplings between the subsystems. Since
the two subsystems are coupled for all nonzero $\Theta$, the
mean-field Hamiltonian at nonzero $\Theta$ can be viewed only as a
Chern insulator with a $C=2$ band. Therefore the topological field
theory approach will predict that the TOS of the system is described
by $SU(2)_2$ Chern-Simons theory coupled to $C=2$ partons, as we
discussed in Sec. \ref{sec2}. In contrast, our numerical result for
small $\Theta$ finds an Abelian TOS, which provides a concrete
example of a case when the TOS of the Gutzwiller-projected wave
function is different from the prediction of topological field
theory.

\subsection{Theoretical interpretation of the topological phase transition} \label{sec3C}

To understand physically the topological phase transition, we first
ask why the derivation of the effective field theory in Sec
\ref{sec2B} does not apply to $\Theta=0$. For general $\Theta$, the
constraints on the partons induces an $SU(2)$ gauge field along all
lattice edges in Fig. \ref{fig7} that dominates the low-energy
theory after the partons are integrated out. In the $\Theta=0$
limit, however, the Hamiltonian becomes Eq. \ref{Ham0}, and all
hoppings between the two subsystems vanish. Therefore there are two
well-defined $SU(2)$ gauge fields in the long wavelength limit, one
for each subsystem. As is clear in Fig. \ref{fig8}, these two
$SU(2)$ gauge fields exist on independent pieces and remain
independent after the partons are integrated out. Integrating out
the $C=1$ band of the parton gives the $SU(2)$ level $1$
Chern-Simons theory, so that the topological field theory of the
$\Theta=0$ system consists of fermions coupling to $SU(2)_1\times
SU(2)_1$.

At finite $\Theta$, coupling is turned on between the two effective
``layers" and breaks the separate $SU(2)\times SU(2)$ gauge symmetry
into one single $SU(2)$. As an alternative view of the symmetry
breaking, one can carry out the unitary rotation in Eq.
\ref{unitrans} in reverse to transform the Hamiltonian $H_\Theta$
back to $H_0$. In the new basis, the partons occupy the two
decoupled $C=1$ bands before projection. The only way the two
independent layers are coupled is through the constraint. In the
original basis the constraint is written as
$n_{iI\uparrow}=n_{iI\downarrow}$
($c_{iI\uparrow}^{\dagger}c_{iI\uparrow}=c_{iI\downarrow}^{\dagger}c_{iI\downarrow}$)
in real space. After the inverse unitary transformation for a finite
$\Theta$, the resulting constraints are
$c_{i1\uparrow}^{\dagger}c_{i1\uparrow}+c_{i1\uparrow}^{\dagger}c_{i1\uparrow}=c_{
i1\downarrow}^{\dagger}c_{i1\downarrow}+c_{i1\downarrow}^{\dagger}c_{i1\downarrow}$
and $\cos\Theta \left( c_{i1\uparrow }^{\dagger}c_{
i1\uparrow}-c_{i1\uparrow }^{\dagger}c_{
i1\uparrow}\right)+\sin\Theta\left(c_{
i2\uparrow}^{\dagger}c_{i1\uparrow }+c_{
i1\uparrow}^{\dagger}c_{i2\uparrow }\right)=\cos\Theta\left(
c_{i1\downarrow }^{\dagger}c_{i1\downarrow }-c_{
i1\downarrow}^{\dagger}c_{ i1\downarrow}\right)+\sin\Theta\left(c_{
i2\downarrow}^{\dagger}c_{i1\downarrow }+c_{
i1\downarrow}^{\dagger}c_{i2\downarrow }\right)$. The latter
explicitly breaks the intra-layer charge conservation symmetry of
the parent Hamiltonian in Eq.\ref{Ham0}, defined by $c_{iIs}^\dagger
\rightarrow e^{-i\phi}c_{iIs}^\dagger,c_{iIs} \rightarrow
e^{i\phi}c_{iIs}$, $x_i+I\in \mbox{odd}$. As a consequence of this
inter-layer coupling, the two $SU(2)$ gauge fields in the effective
theory are coupled and only a diagonal $SU(2)$ gauge symmetry is
preserved. Physically, the holes in the two $C=1$ bands are no
longer distinguishable so that the two semionic quasiparticles
originating from the holes in the two bands now merge to one
particle. Consequently, the ground-state degeneracy on a torus,
effectively labeled by the quasiparticle content, also decreases
from four fold to three fold.

\begin{figure}
\begin{center}
\includegraphics[scale=0.4]{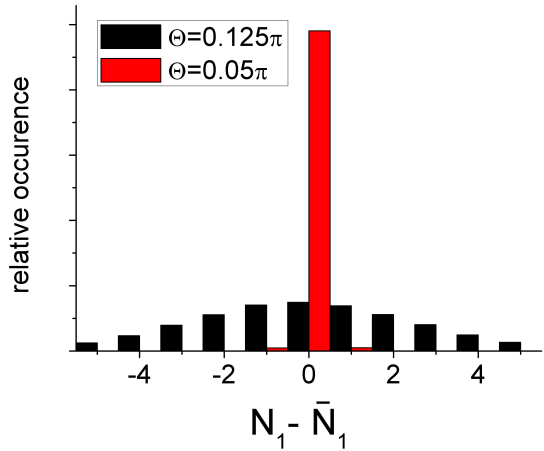}
\caption{A histogram of the number of sampled configurations versus
the parton number around its average $N_1- \bar N_1$ in one of the
decomposed $C=1$ bands. While the $N_1=\bar N_1$ central peak
contains more than $98\%$ of the configurations for $\Theta=0.05\pi$
(red), the spread for $\Theta=0.125\pi$ (black) is much wider and
the percentage of the $N_1=\bar N_1$ configurations is only $15\%$
suggesting that $N_1$ is no longer a good quantum number. The
results are obtained on system size $L_x=L_y=28$ with periodic
boundary conditions.} \label{fig5}
\end{center}
\end{figure}

\begin{figure}
\begin{center}
\includegraphics[scale=0.4]{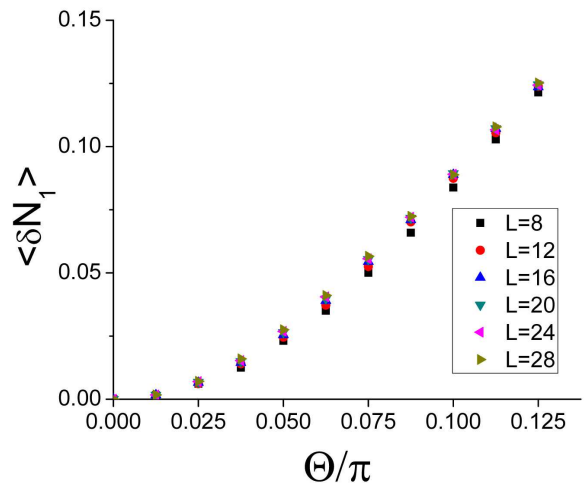}
\caption{The mean squared deviation $\sqrt{\left\langle
\left(N_1-\bar N_1\right)^2\right\rangle/\bar N_1}$ versus $\Theta$
for system sizes $L_x=L_y=8,12,16,20,24,28$.} \label{fig6}
\end{center}
\end{figure}

The discussion above suggests that the Abelian and non-Abelian
phases are distinguished by whether the two layers (in the rotated
parton basis) have separately conserved particle numbers. In the
Abelian (non-Abelian) phase, the separate particle number
conservation of the two layers is effective preserved (broken). To
verify this scenario, we numerically calculate the fluctuations of
parton number in one of the $C=1$ layers (in the rotated parton
basis): $N_1=\underset{I+x_i \in \mbox{odd}}{\sum} c^\dagger_{
iI\uparrow} c_{iI\uparrow}$. In the $\Theta=0$ limit, the two bands
are independent, therefore $N_1=\bar N_1$ and the fluctuation is
exactly zero. As $\Theta$ increases, the intra-band charge
conservation is broken, and therefore one may expect an increase in
the $N_1$ fluctuation. Fig. \ref{fig5} is a histogram of the number
of sampled configurations in the projected wave function versus the
parton number $N_1$ fluctuation around its average value $\bar N_1$
in one of the $C=1$ bands at $\Theta=\pi/20$ (red) and
$\Theta=\pi/8$ (black). While such fluctuation is still largely
suppressed and the $N_1$ conservation approximately holds at
$\Theta=\pi/20$ on the Abelian TOS side of the transition, it
proliferates at $\Theta=\pi/8$ and the intralayer charge
conservation no longer exists for a non-Abelian TOS. To see further
the connection between the parton number fluctuation and the
non-Abelian TOS, we show in Fig. \ref{fig6} the mean squared
deviation $\sqrt{\left\langle \left(N_1-\bar
N_1\right)^2\right\rangle/\bar N_1}$ versus $\Theta$ for various
system sizes.

In reality, for a multiband TOS such as the topological nematic
states\cite{Maissam2012}, band mixing, be it hopping or interaction,
is hard to eliminate. The existence of a finite $\Theta_c$ suggests
that the Abelian TOS is stable against weak band-mixing
perturbations. Intuitively, this is because the TOS are protected by
excitation gaps. For small band-mixing perturbations, the charge
conservation within the bands can appear as an emergent symmetry.
Nevertheless, in comparison with integer Chern insulators protected
by the band gap, the TOS are relatively vulnerable. A topological
phase transition can occur even if the band structure remains
adiabatically equivalent.

\section{Conclusions}\label{sec4}

In conclusion, we study topological properties of non-Abelian TOS
using Gutzwiller-projected wave functions and the momentum
polarization approach. Our numerical results on the topological spin
and edge central charge confirm that projected wave functions of two
partons in Chern bands with Chern number $C=2$ are described by the
field theory of $\nu=2$ fermions coupled to an $SU(2)$ gauge field,
and clearly distinguish it from the pure $SU(2)_2$ CS theory. In
addition, we adiabatically interpolate the parent Chern insulator
with $C=2$ with two Chern insulators each with $C=1$, and track the
variation of topological quantities such as the topological spin and
ground-state degeneracy for their corresponding TOS projected wave
functions. We show that the topological phase transition between the
non-Abelian and Abelian TOS is marked by the breaking down of charge
conservation within each of the $C=1$ Chern bands. The transition
point is close to but apart from the completely decoupled limit, in
consistency with the intuition that the corresponding Abelian TOS is
protected by a gap and stable against small band-mixing
perturbations. Our result demonstrates explicitly that the
topological order in a Gutzwiller-projected state does not always
agree with the prediction of topological field theory, and
generically has to be determined by numerical calculations of
topological properties.

Our numerical methods based on momentum polarization and the
variational Monte Carlo method are generalizable to more complicated
non-Abelian TOSs described by Gutzwiller-projected wave functions.
Compared to previous approaches, momentum polarization provides an
efficient way to extract characteristic quantities given the
many-body wave functions of a chiral topological ordered state. One
open question left for future work is whether the critical behavior
of momentum polarization across a topological phase transition can
be studied numerically and compared with any field theory
description. Another open question is whether there is a more
generic proof of the momentum polarization formula in Eq.
\ref{Mompol}, which has been verified numerically in several TOS,
but has not been proved analytically except for arguments based on
edge-state conformal field theory\cite{mompol}.

We would like to thank Maissam Barkeshli, Chao-Ming Jian, Ching Hua
Lee and Peng Ye for insightful discussions. This work is supported
by the Stanford Institute for Theoretical Physics (YZ) and the
National Science Foundation through the grant No. DMR-1151786 (XLQ).

\def\urlprefix{}
\def\url#1{}
\bibliographystyle{apsrev}
\bibliography{bibliography}

\end{document}